\documentstyle[12pt]{article}
\newcommand{\beq}{\begin{equation}}
\newcommand{\eeq}{\end{equation}}
\newcommand{\ba}{\begin{eqnarray}}
\newcommand{\ea}{\end{eqnarray}}

\newcommand{\dsl}
  {\kern.06em\hbox{\raise.15ex\hbox{$/$}\kern-.56em\hbox{$\partial$}}}

\newcommand{\eeqarr}{\end{eqnarray}}
\newcommand{\ZZ}{{\rm \kern 0.275em Z \kern -0.92em Z}\;}
\begin{document}
\begin{titlepage}
\begin{center}
{\Huge Scalar Field Theory in the}
\\
\vspace*{0.5cm}
{\Huge AdS/CFT Correspondence Revisited}
\\
\vspace*{0.5cm}
{\large Pablo Minces\footnote{pablo@fma.if.usp.br} \\ and \\ 
Victor O. Rivelles\footnote{rivelles@fma.if.usp.br}}
\\
\vspace*{0.5cm}
Universidade de S\~ao Paulo, Instituto de F\'{\i}sica\\
Caixa Postal 66.318 - CEP 05315-970 - S\~ao Paulo - Brazil
\vspace*{0.5cm}
\end{center}

\begin{abstract}
We consider the role of boundary conditions in the $AdS_{d+1}/CFT_{d}$
correspondence for the scalar field theory. Also a careful analysis of
some limiting cases is presented. 
We study three possible types of boundary conditions, Dirichlet,
Neumann and mixed. We compute the two-point functions of the conformal
operators on the boundary for each type of boundary condition. We show
how particular choices of the mass require different treatments. In the
Dirichlet case we find that there is no double zero in the two-point
function of the operator with conformal dimension $\frac{d}{2}$. The
Neumann case leads to new normalizations for the boundary two-point
functions. In the
massless case we show that the conformal dimension
of the boundary conformal operator is precisely the unitarity bound for
scalar operators. We find a one-parameter family
of boundary conditions in the mixed case. There are   
again new normalizations for the boundary two-point
functions. For a particular choice of the mixed boundary condition and
with the mass squared in the range $-d^2/4<m^2<-d^2/4+1$ the
boundary operator has conformal dimension comprised in the interval
$\left[\frac{d-2}{2}\; ,\;\frac{d}{2}\right]$. For mass squared
$m^2>-d^2/4+1$ the same choice of mixed boundary condition leads to a
boundary operator whose conformal dimension is the unitarity bound.
\end{abstract}
\vspace*{0.5cm}
\begin{flushleft}
\end{flushleft}
\end{titlepage}

\section{Introduction}

Since the proposal of Maldacena's conjecture, which gives a
correspondence 
between a field theory on anti-de Sitter space (AdS) and a conformal
field theory (CFT) on its boundary \cite{malda}, an intensive work has
been devoted to get a deeper understanding of its implications. 
In particular, a precise form to the conjecture has been given in
\cite{polyakov}\cite{witten}. It reads
\beq
Z _{AdS}[\phi _{0}] = \int _{\phi _{0}}{\cal D}\phi\
\exp\left(-I[\phi]\right)\equiv Z _{CFT}[\phi _{0}] = \left<\exp\left(\int
_{\partial\Omega}d^{d}x{\cal O}\phi _{0}\right)\right>\; ,
\label{1}
\eeq
where $\phi _{0}$ is the boundary value of the bulk field $\phi$ which
couples to the boundary CFT operator ${\cal O}$. This allows us to obtain 
the correlation functions of the boundary CFT theory in d dimensions by
calculating the partition function on the $AdS _{d+1}$ side.
The AdS/CFT correspondence has been studied for the scalar field
\cite{witten}\cite{viswa2}\cite{freedman}\cite{kraus},
the vector field \cite{witten}\cite{freedman}\cite{viswa}\cite{our}, 
the spinor field \cite{viswa}\cite{sfetsos}\cite{kaviani},
the Rarita-Schwinger field \cite{volovich}\cite{koshelev}\cite{viswa4},
the graviton
field \cite{liu}\cite{viswa5}, the massive symmetric tensor field
\cite{polishchuk}
and the antisymmetric $p$-form field \cite{frolov}\cite{l'yi}. In all
cases
Dirichlet boundary conditions were used. Several subtle points have been
clarified in these papers and all results lend support to the conjecture. 

In a broader sense Maldacena's conjecture is a concrete realization of
the holographic principle \cite{susskind}\cite{hooft}. We expect that any 
field theory relationship in AdS space must be reflected in the border
CFT. An example of this is the well known equivalence between
Maxwell-Chern-Simons theory and the self-dual model in three
dimensional Minkowski space \cite{equivalence}. This equivalence
holds also in $AdS_3$ and using the AdS/CFT correspondence we have shown
that the corresponding boundary operators have the same conformal
dimensions \cite{our}. Another situation involves massive scalar fields in
AdS spaces. If the scalar field  has mass-squared in the
range $-d^2/4<m^2<-d^2/4+1$ then there are two possible quantum field
theories in the bulk \cite{freedman2}. The AdS/CFT correspondence with
Dirichlet boundary condition can easily
account for one of the theories. The other one appears in a very
subtle way by identifying a conjugate field through a Legendre
transform as the source of the boundary conformal operator \cite{witten2}.
The existence of two conjugated boundary operators has been first pointed
out in \cite{dobrev}.  

Since a field theory is determined not only by its
Lagrangian but also by its boundary terms in the action we expect
that the AdS/CFT correspondence must be sensitive to these 
boundary terms. This is easily seen to be true by computing the
left-hand side of Eq.(\ref{1}) for a classical field
configuration. All that is left is a boundary term. If we start with
different boundary terms in the action then we obtain different
correlation functions on the right-hand side. 

The origin of boundary terms in the action is due to the variational
principle. In order to have a stationary action boundary terms,
which will depend on the choice of the boundary conditions, must be
introduced. The importance of these
boundary terms for the AdS/CFT correspondence was recognized in the
case of spinor fields where the 
action is of first order in derivatives and the classical action
vanishes on-shell \cite{henneaux}. They also played an important role
in the case of Chern-Simons theory \cite{our}. Therefore it is crucial
to understand the implications of different types of 
boundary conditions for the same theory since they in general imply
different boundary terms. 

In this work we will study the role of
different types of boundary condition for the scalar field theory. 
We will consider Dirichlet and Neumann boundary conditions, and a
combination of both of them which we will call mixed boundary condition.
Each type of boundary condition requires a different boundary term. We
will show that the mixed boundary conditions are parametrized by a real
number so that there is a one-parameter family of boundary terms
consistent with the variational principle. 

We will also show that different types of boundary condition give
rise to different conformal field theories at the border. For the scalar
field this was somehow expected. The two solutions found in
\cite{freedman2} correspond to two different choices of 
energy-momentum tensor. Both of them are conserved and their
difference gives a surface contribution to the isometry
generators. Although these two solutions were found in the Hamiltonian
context by requiring finiteness of the energy they will reappear here
by considering different types of boundary condition which amounts to
different boundary terms in the action. We can also
look for the asymptotic behavior of the scalar field near the
boundary according to the chosen type of boundary condition. 
For the Dirichlet boundary condition it is well known that
the scalar field behaves as $x_0^{d/2 - \sqrt{d^2/4+m^2}}$
near the border at $x_0=0$. There is no upper restriction on the mass in
this case. It corresponds to one of the solutions found in
\cite{freedman2} and gives rise to a boundary conformal operator
with conformal dimension  $d/2 + \sqrt{d^2/4+m^2}$. We will show that for
a particular
choice of mixed boundary condition and when the mass squared is in the
range $-d^2/4 < m^2 < -d^2/4+1 $ the scalar field behaves as
$x_0^{d/2 + \sqrt{d^2/4+m^2}}$ near the border. It corresponds
precisely to the second solution of 
\cite{freedman2} and gives rise to a boundary conformal operator
with conformal dimension  $d/2 - \sqrt{d^2/4+m^2}$. Note that the upper
limit for the mass squared $-d^2/4+1$ is consistent with the unitarity
bound $(d-2)/2$.

Another important point that we will show is the existence of boundary
conditions which give rise to boundary conformal operators for which
the unitarity bound $(d-2)/2$ is 
reached. They correspond to a massless scalar field with Neumann
boundary condition or to a massive scalar field with $m^2 > -d^2/4 +1$
with a
particular choice of the mixed boundary condition (the same choice
which gives the boundary operator with conformal dimension
$d/2-\sqrt{d^2/4+m^2}$). In this way, using different boundary
conditions, we obtain all scalar conformal
field theories allowed by the unitarity bound. 

We will also analyze carefully two cases where the mass of the scalar
field takes special values. In some cases the usual expansion of the
modified Bessel functions in powers of $x_0$ breaks down and we must
use expansions involving logarithms. When $ m^2=-d^2/4$ it gives rise
to the asymptotic behavior $x_0^{d/2} \ln x_0$  and the two-point
function is obtained without troubles. This is to be contrasted with
the usual limiting 
procedure where the mass goes to $ m^2=-d^2/4$ but the two-point
function has a double zero in the limit \cite{witten2}. The other case
corresponds to $\sqrt{d^2/4+m^2}$ integer but non-zero. In this case we
just reproduce the known results. 

We should stress the fact that the use of different types of boundary
condition (for given values of $m^2$ and $d$) allows us in general to get
boundary two-point functions with different normalizations. This will
affect the
three-point and higher-point functions. Maybe this is related
to
the fact that AdS and field theory calculations agree up to some
dimension dependent normalization factors \cite{lee} but we will not
discuss this further.  

The paper is organized as follows. In section 2 we find the
boundary
terms corresponding to each type of boundary condition. In section
3 we consider the Dirichlet case while in Section 4 we treat Neumann
boundary conditions. In section 5 we consider mixed boundary conditions.
Finally section 6 presents our conclusions.
In appendix A we list all boundary two-point
functions that
we computed and appendix B contains some useful formulae.

\section{The Variational Principle}

We take the usual Euclidean representation of the $AdS _{d+1}$
in Poincar\'e coordinates described by the 
half space $x _{0}>0$, $x _{i} \in {\bf R}$ with metric
\beq
ds^{2}=\frac{1}{x _{0}^{2}} \sum_{\mu=0}^{d} dx^{\mu}dx^{\mu}.
\label{2}
\eeq
The action for the massive scalar field theory is given by 
\beq
I_{0} =  \frac{1}{2}\int d^{d+1}x \, \sqrt{g} \left(
g^{\mu\nu} \partial_{\mu}
\phi\partial_{\nu}\phi\;+\;m^{2}\phi^{2}\right),
\label{3}
\eeq 
and the corresponding equation of motion is
\beq
\left( \nabla^2 - m^2\right)\phi=0.
\label{4}
\eeq
The solution which is regular at $x_0 \rightarrow \infty$ reads
\cite{freedman}
\beq
\phi(x) = \int\frac{d^{d}k}{\left(2\pi\right)^{d}}\;
e^{-i\vec{k}\cdot\vec{x}}\;
x_{0}^{\frac{d}{2}}\;a(\vec{k})\;K_{\nu}(kx_{0}),
\label{4'}
\eeq
where $\vec{x}=(x^{1},...,x^{d})$, $k = \mid\vec{k}\mid$,
$K_{\nu}$ is the modified Bessel function, and
\beq
\nu = \sqrt{\frac{d^{2}}{4}+m^{2}}\; .
\label{5}
\eeq
From Eq.(\ref{4'}) we also get
\beq
\partial_{0}\phi(x) = \int\frac{d^{d}k}{\left(2\pi\right)^{d}}\;
e^{-i\vec{k}\cdot\vec{x}}\;
x_{0}^{\frac{d}{2}-1}\;a(\vec{k})\;\left[\left(\frac{d}{2} + \nu
\right)K_{\nu}(kx_{0}) - kx_{0}K_{\nu + 1}(kx_{0})\right].
\label{5'}
\eeq

In order to have a stationary action we must supplement the action 
$I_0$ with a boundary term $I_S$ which cancels its variation. The
appropriate action is then 
\beq
I=I_{0} + I_{S}.
\label{6}
\eeq
In order to capture the effect of the Minkowski boundary of the
$AdS_{d+1}$, situated at $x_0=0$, we first consider a boundary
value problem on the boundary surface $x_0=\epsilon > 0$ and
then take the limit $\epsilon \rightarrow 0$ at the very end. Then the
variational principle applied to the action $I$ gives 
\beq
\delta I = -\int d^{d}x \;\epsilon^{-d+1} 
\;\partial_{0}\phi_{\epsilon}\;\delta\phi_{\epsilon} + \delta
I_{S} = 0, 
\label{8}
\eeq
where $\phi_\epsilon$ and $\partial_{0}\phi_{\epsilon}$ are the value
of the field and its derivate at $x_0=\epsilon$ respectively. 
This equation will be used below to find out the appropriate
boundary term $I_S$ for each type of boundary condition.

For Dirichlet boundary condition the variation of the field at the
border vanishes so that the first term in Eq.(\ref{8}) also vanishes
and the usual action $I_{0}$ is already stationary.  
Making use of the field equation the action $I$ takes the form 
\beq
I_{D}= \frac{1}{2}\int
d^{d+1}x \; \partial_{\mu}\left(
\sqrt{g}\;\phi\;\partial^{\mu}\phi\right)=-\frac{1}{2}\int d^{d}x
\;\epsilon^{-d+1}
\;\phi_{\epsilon}\;\partial_{0}\phi_{\epsilon}\;.
\label{11}
\eeq
It is to be understood that
$\partial_{0}\phi_{\epsilon}$ in Eq.(\ref{11}) is evaluated in terms
of the Dirichlet data $\phi_{\epsilon}$. 

To consider Neumann boundary conditions we first take a unitary
vector which is inward normal to the boundary $
n^{\mu}(x_{0})=(x_{0},{\bf 0})$. The Neumann boundary condition then
fixes the value of $n^{\mu}(\epsilon) \partial_{\mu}\phi_\epsilon
\equiv \partial_n \phi_\epsilon$. The boundary term to be added to the
action reads
\beq
I_S= - \int
d^{d+1}x \; \partial_{\mu}\left(
\sqrt{g}\; g^{\mu\nu} \; \phi\;\partial_{\nu}\phi\right)=\int d^{d}x
\;\epsilon^{-d+1} 
\;\phi_{\epsilon}\;\partial_{0}\phi_{\epsilon}\;,
\label{9'}
\eeq
so that we find the following
expression for the action at the boundary
\beq
I_{N}=
\frac{1}{2}\int d^{d}x
\;\epsilon^{-d}
\;\phi_{\epsilon}\;\partial_{n}\phi_{\epsilon}\;.
\label{12}
\eeq
Here $\phi_{\epsilon}$ is to be expressed in terms of the Neumann
value $\partial_{n}\phi_{\epsilon}$. 
Notice that the 
on-shell value of the action with Neumann boundary condition
Eq.(\ref{12}) differs by a sign from the corresponding action with
Dirichlet boundary condition Eq.(\ref{11}). 

We now consider a boundary condition which fixes the value of a
linear combination of the field and its normal derivative at the border
\beq
\phi(x) + \alpha n^{\mu}\partial_{\mu}\phi(x) \equiv \psi^{\alpha}(x)
\; . 
\label{100}
\eeq
We will call it as mixed boundary condition. Here $\alpha$ is an arbitrary 
real but non-zero
coefficient. In this case the surface term to be added to the action
is 
\beq
I_{S}^{\alpha} = \frac{\alpha}{2}\int
d^{d+1}x \; \partial_{\mu}\left(
\sqrt{g}\; g^{\mu\nu} \partial_{\nu}\phi\; n^\rho \partial_{\rho}\phi
\right)=
-\frac{\alpha}{2}\int d^{d}x\;\epsilon^{-d+2}
\;\partial_{0}\phi_{\epsilon}\;\partial_{0}\phi_{\epsilon}\;,
\label{102}
\eeq
and we find the following expression for the action at the boundary
\beq
I_{M}^{\alpha} = -\frac{1}{2}\int d^{d}x
\;\epsilon^{-d+1}
\;\psi^{\alpha}_{\epsilon}\;\partial_{0}\phi_{\epsilon}\;.
\label{104}
\eeq
Clearly $\partial_{0}\phi_{\epsilon}$ in the above expression must be
written in terms of the boundary data $\psi^{\alpha}_{\epsilon}$. 
We then have a one-parameter family of surface terms since the
variational 
principle does not impose any condition on $\alpha$. In this way the
value of the on-shell action Eq.(\ref{104}) also depends on $\alpha$. 

In the following sections we will consider each boundary condition
separately.

\section{Dirichlet Boundary Condition}

We begin by recalling the main results for the Dirichlet case
\cite{viswa2}\cite{freedman}. Let $\phi_{\epsilon}(\vec{k})$ be the
Fourier transform
of the Dirichlet boundary value of the field $\phi(x)$. From
Eq.(\ref{4'}) we get 
\beq
a(\vec{k})=\frac{\epsilon^{-\frac{d}{2}}\;\phi_{\epsilon}(\vec{k})}
{K_{\nu}(k\epsilon)}\; ,
\label{13}
\eeq
and inserting this into Eq.(\ref{5'}) we find
\beq
\partial_{0}\phi_{\epsilon}(\vec{x})=\int
d^{d}y\;\phi_{\epsilon}(\vec{y})\int\frac{d^{d}k}{\left(
2\pi\right)^{d}}\;e^{-i\vec{k}\cdot\left(
\vec{x}-\vec{y}\right)}\;\epsilon^{-1}\;\left[\frac{d}{2} + \nu -
k\epsilon\;\frac{K_{\nu+1}(k\epsilon)}{K_{\nu}(k\epsilon)}\right].
\label{14}
\eeq
Then the action Eq.(\ref{11}) reads
\beq
I_{D}=-\frac{1}{2}\int d^{d}x \; d^{d}y\;\phi_{\epsilon}(\vec{x})
\;\phi_{\epsilon}(\vec{y})\;\epsilon^{-d}\int\frac{d^{d}k}{\left(
2\pi\right)^{d}}\;e^{-i\vec{k}\cdot\left(
\vec{x}-\vec{y}\right)}\;\left[\frac{d}{2} + \nu -
k\epsilon\;\frac{K_{\nu+1}(k\epsilon)}{K_{\nu}(k\epsilon)}\right].
\label{15}
\eeq

The next step is to keep the relevant terms in the series expansions of
the Bessel functions and to integrate in $\vec{k}$. We consider first
the case $\nu\not=0$ that is $m^2 \not= -\frac{d^2}{4}$. For completeness
we
list the relevant modified Bessel functions in Appendix B. For 
$\nu$ not integer we make use of Eqs.(\ref{16},\ref{16'}), whereas 
for $\nu$ integer but non-zero we use
Eqs.(\ref{16'''},\ref{16'''''}). In  both cases we get the same result 
\beq
I_{D}^{\nu\not=0}=-\frac{\nu}{\pi^{\frac{d}{2}}}\;\frac{\Gamma(\frac{d}{2}
+ \nu)}{\Gamma(\nu)}
\; \int d^{d}x \;
d^{d}y\;\phi_{\epsilon}(\vec{x})
\;\phi_{\epsilon}(\vec{y})\;\frac{\epsilon^{2(\nu-\frac{d}{2})}}{|\;
\vec{x}-\vec{y}\;|^{2(\frac{d}{2}+\nu)}}\;+\;\cdots,
\label{17}
\eeq
where the dots stand for either contact terms or higher order terms in
$\epsilon$. 

Taking the limit \cite{freedman}
\beq
\lim _{\epsilon\rightarrow
0}\epsilon^{\nu-\frac{d}{2}}\phi_{\epsilon}(\vec{x})
= \phi_{0}(\vec{x}),
\label{18}
\eeq
to go to the border and making use of the AdS/CFT equivalence in the form
\beq
\exp\left( -I _{AdS}\right) \equiv \left<\exp\left(\int d^{d}x \;
{\cal O}(\vec{x}) \; \phi_{0}(\vec{x})\right)\right>,
\label{19}
\eeq
we find the following two-point function
\beq
\left<{\cal O}_{D}^{\nu\not=0}(\vec{x}){\cal
O}_{D}^{\nu\not=0}(\vec{y})\right>=\frac{2\nu}{\pi^{\frac{d}{2}}}\;
\frac{\Gamma(\frac{d}{2}+\nu)}{\Gamma(\nu)}\;\frac{1}{|\;
\vec{x}-\vec{y}\;|^{2(\frac{d}{2}+\nu)}}\; .
\label{20}
\eeq
Then the conformal operator ${\cal O}_{D}^{\nu\not=0}$ on the boundary
CFT has conformal dimension $\frac{d}{2}+\nu$. From Eq.(\ref{18}) we
find that near the border $\phi$ behaves as $x_0^{d/2-\nu}
\phi_0(\vec{x})$ as expected. In this way we have
extended the results of \cite{viswa2}\cite{freedman} to the case
$\nu$ integer but non-zero. 

For future reference we note that in the particular case $m=0$, that
is $\nu=\frac{d}{2}$, Eq.(\ref{20}) reads
\beq
\left<{\cal O}_{D}^{\nu=\frac{d}{2}}(\vec{x}){\cal
O}_{D}^{\nu=\frac{d}{2}}(\vec{y})\right>=\frac{d}{\pi^{\frac{d}{2}}}\;
\frac{\Gamma(d)}{\Gamma(\frac{d}{2})}\;\frac{1}{|\;
\vec{x}-\vec{y}\;|^{2d}}\;,
\label{21}
\eeq
so that the operator ${\cal O}_{D}^{\nu=\frac{d}{2}}$ has conformal
dimension $d$.

Now we consider the case $\nu=0$, that is
$m^{2}=-\frac{d^{2}}{4}$. Since the two-point function  
Eq.(\ref{20}) has a double zero for $\nu=0$ it was  
argued \cite{witten2} that the correct result can be found by introducing a
normalization on the boundary operator. Instead we will make use of the
expansion for the Bessel function $K_{0}$. 
Using Eqs.(\ref{16'''},\ref{827}) we get
\ba
k\epsilon\;\frac{K_{1}(k\epsilon)}{K_{0}(k\epsilon)}&=&
-\frac{1}{ln\;\epsilon}\left[ 1\;
+\;\frac{(k\epsilon)^{2}}{2}\;ln\;\epsilon\; +\;
O\left(\epsilon^{2}\right)\right]\left[1-\frac{ln\;
k\;-\;ln\; 2\;+\;\gamma}{ln\;\epsilon}\; +\;
O\left(\epsilon^{2}\right) \right]\nonumber\\&=&\frac{ln\;
k}{ln^{2}\epsilon}\; +\; \cdots,
\label{3000} 
\ea
where the dots denote all other terms representing either contact
terms in the two-point function or terms of higher order in $\epsilon$.
Notice that it is essential to separate the contributions of $k$ and
$\epsilon$ in the terms $ln\; k\epsilon$ in order to identify the relevant
contributions. Substituting in Eq.(\ref{15}) we find
\beq
I_{D}^{\nu=0}=
\frac{1}{2}\int
d^{d}x\; d^{d}y\;\phi_{\epsilon}(\vec{x})
\;\phi_{\epsilon}(\vec{y})\;\frac{\epsilon^{-d}}{ln^{2}\epsilon}\;
\int\frac{d^{d}k}{\left(
2\pi\right)^{d}}\; e^{-i\vec{k}\cdot 
(\vec{x}-\vec{y})}\;ln\; k\;+\;\cdots.
\label{3001}
\eeq 
The integration in $\vec{k}$ is carried out by making use of
Eq.(\ref{16'''''}) yielding 
\beq
I_{D}^{\nu=0}=-\frac{\Gamma\left(
\frac{d}{2}
\right)}{4\pi^{\frac{d}{2}}}\; \int d^{d}x \;
d^{d}y\;\phi_{\epsilon}(\vec{x})
\;\phi_{\epsilon}(\vec{y})\;\frac{\epsilon^{-d}}{ln^{2}\epsilon}\;
\frac{1}{|\;
\vec{x}-\vec{y}\;|^{d}}\;+\;\cdots .
\label{466}
\eeq
To go to the border we have to rescale $\phi_\epsilon$ using a factor
of $\ln \epsilon$. This makes the rescaling somewhat arbitrary since
any power of $\epsilon$ in $\ln \epsilon$ would do the job. So
choosing the limit 
\beq
\lim _{\epsilon\rightarrow
0}\; (\epsilon^{\frac{d}{2}} \; ln\;\epsilon)^{-1} \;
\phi_{\epsilon}(\vec{x}) = \phi_{0}(\vec{x}), 
\label{467}
\eeq
and making use of the AdS/CFT equivalence Eq.(\ref{19}) we find the
following two-point function
\beq
\left<{\cal O}_{D}^{\nu=0}(\vec{x}){\cal   
O}_{D}^{\nu=0}(\vec{y})\right>=
\frac{\Gamma\left(\frac{d}{2}\right)}{2\pi^{\frac{d}{2}}}\;
\frac{1}{|\;
\vec{x}-\vec{y}\;|^{d}}\; .
\label{128}
\eeq
Then the conformal operator ${\cal O}_{D}^{\nu=0}$ on the boundary
CFT has conformal dimension $\frac{d}{2}$ as expected. As anticipated
in \cite{freedman} the scalar field approaches the boundary as
$x_0^{d/2}\ln x_0\;\phi_{0}(\vec{x})$ due to the logarithm appearing in
the expansion of
the Bessel function. 

\section{Neumann Boundary Condition}

Using the Neumann boundary condition we get from Eq.(\ref{5'}) 
\beq
a(\vec{k})=\frac{\epsilon^{-\frac{d}{2}}\;\partial_{n}
\phi_{\epsilon}(\vec{k})}
{(\frac{d}{2}+\nu)K_{\nu}(k\epsilon)-k\epsilon K_{\nu+1}(k\epsilon)}\; ,
\label{22}
\eeq
and substituting this in Eq.(\ref{4'}) we find
\beq
\phi_{\epsilon}(\vec{x})=\int
d^{d}y\;\partial_{n}\phi_{\epsilon}(\vec{y})\int\frac{d^{d}k}{\left(
2\pi\right)^{d}}\;e^{-i\vec{k}\cdot\left(
\vec{x}-\vec{y}\right)}\;\frac{1}{\frac{d}{2}+\nu-
k\epsilon\;\frac{K_{\nu+1}(k\epsilon)}{K_{\nu}(k\epsilon)}}\; .
\label{23}
\eeq
Then the action Eq.(\ref{12}) reads
\beq
I_{N}=\frac{1}{2}\int d^{d}x \;
d^{d}y\;\partial_{n}\phi_{\epsilon}(\vec{x})
\;\partial_{n}\phi_{\epsilon}(\vec{y})\;\epsilon^{-d}\int
\frac{d^{d}k}{\left( 2\pi\right)^{d}}\;e^{-i\vec{k}\cdot\left(
\vec{x}-\vec{y}\right)}\;\frac{1}{\frac{d}{2}+\nu-
k\epsilon\;\frac{K_{\nu+1}(k\epsilon)}{K_{\nu}(k\epsilon)}}\;.
\label{24}
\eeq
In order to keep the relevant terms in the series expansions of the Bessel
functions we must consider the massive and massless cases separately. 

In the massless case we have $\nu=\frac{d}{2}$. For $d$ odd we make
use of Eq.(\ref{16}), whereas for $d$ even we use 
Eq.(\ref{16'''}). In both cases we get for $d>2$ 
\beq
\frac{1}{\frac{d}{2} + \nu -
k\epsilon\;\frac{K_{\nu+1}(k\epsilon)}{K_{\nu}(k\epsilon)}}=
-(d-2)(k\epsilon)^{-2}\; +\;\cdots,
\label{3002}
\eeq
up to contact terms and higher order terms in $\epsilon$. 
Substituting this in Eq.(\ref{24}) we find
\beq
I_{N}^{\nu=\frac{d}{2}}=
-\frac{d-2}{2}\int
d^{d}x\; d^{d}y\;\partial_{n}\phi_{\epsilon}(\vec{x})
\;\partial_{n}\phi_{\epsilon}(\vec{y})\;
{\epsilon^{-d-2}}\;
\int\frac{d^{d}k}{\left(
2\pi\right)^{d}}\; e^{-i\vec{k}\cdot
(\vec{x}-\vec{y})}\;k^{-2}\;+\;\cdots,
\label{3003}
\eeq
and performing the integral in $\vec{k}$ we get 
\beq
I_{N}^{\nu=\frac{d}{2}}=-\frac{\Gamma(\frac{d}{2})}{4\pi^{\frac{d}{2}}}\;
\int d^{d}x \;
d^{d}y\; \partial_{n}\phi_{\epsilon}(\vec{x}) \;
\partial_{n}\phi_{\epsilon}(\vec{y})\; 
\frac{\epsilon^{-d-2}}{|\;
\vec{x}-\vec{y}\;|^{2\frac{d-2}{2}}}\;+\;\cdots,
\label{29}
\eeq
where the dots stand for either contact terms or higher order terms in
$\epsilon$. 

Taking the limit
\beq
\lim _{\epsilon\rightarrow
0}\epsilon^{-\frac{d}{2}-1}\; \partial_{n}\phi_{\epsilon}(\vec{x})
= \partial_{n}\phi_{0}(\vec{x}),
\label{30}
\eeq
and making use of the AdS/CFT equivalence of the form
\beq
\exp\left( -I _{AdS}\right) \equiv \left<\exp\left(\int d^{d}x \;
{\cal O}(\vec{x}) \; \partial_{n}\phi_{0}(\vec{x})\right)\right>,
\label{27}
\eeq
we find the following boundary two-point function
\beq
\left<{\cal O}_{N}^{\nu=\frac{d}{2}}(\vec{x}){\cal
O}_{N}^{\nu=\frac{d}{2}}(\vec{y})\right>=
\frac{\Gamma(\frac{d}{2})}{2\pi^{\frac{d}{2}}}\;\frac{1}{|\;
\vec{x}-\vec{y}\;|^{2\frac{d-2}{2}}}\; .
\label{31}
\eeq
Then for $d>2$, even or odd, the conformal dimension of the operator
${\cal O}_{N}^{\nu=\frac{d}{2}}$ is precisely the unitarity bound
$\frac{d-2}{2}$. From Eq.(\ref{30}) we find that near the border the
scalar field goes as $x_0^{d/2+1}\partial_{n}\phi_0(\vec{x})$.  
Comparing Eqs.(\ref{21},\ref{31}) we see that
the conformal dimensions of the boundary operators for
the massless Dirichlet and Neumann cases are different and for the
later case the unitarity bound is reached. 

For the massive scalar field, that is $\nu\not=\frac{d}{2}$, we first
consider the case $\nu\not=0$ i.e. $m^2 \not= -\frac {d^{2}}{4}$. We have
again
to consider separately the 
cases with $\nu$ not integer and $\nu$ integer but non-zero. In both
cases we find 
\beq
I_{N}^{\nu\not=0,\frac{d}{2}}=-\frac{\nu}{\pi^{\frac{d}{2}}}\;
\frac{1}{\left( \frac{d}{2}-\nu\right)^{2}}\;\frac{\Gamma(\frac{d}{2} 
+ \nu)}{\Gamma(\nu)}\;\int d^{d}x \;
d^{d}y\;\partial_{n}\phi_{\epsilon}(\vec{x})
\;\partial_{n}\phi_{\epsilon}(\vec{y})\;
\frac{\epsilon^{2(\nu-\frac{d}{2})}}{|\;
\vec{x}-\vec{y}\;|^{2(\frac{d}{2}+\nu )}}\; +\cdots.
\label{25}
\eeq
Taking the limit
\beq
\lim _{\epsilon\rightarrow
0}\epsilon^{\nu-\frac{d}{2}}\;\partial_{n}\phi_{\epsilon}(\vec{x})
= \partial_{n}\phi_{0}(\vec{x}),
\label{26}
\eeq
and making use of the AdS/CFT equivalence Eq.(\ref{27}) we find the
following boundary two-point function
\beq
\left<{\cal O}_{N}^{\nu\not=0,\frac{d}{2}}(\vec{x}){\cal
O}_{N}^{\nu\not=0,\frac{d}{2}}(\vec{y})\right>=
\frac{2\nu}{\pi^{\frac{d}{2}}}\;
\frac{1}{\left( \frac{d}{2}-\nu\right)^{2}}\;
\frac{\Gamma(\frac{d}{2}+\nu)}{\Gamma(\nu)}\;\frac{1}{|\;
\vec{x}-\vec{y}\;|^{2(\frac{d}{2}+\nu )}}\; .
\label{28}
\eeq 
Then the operator ${\cal O}_{N}^{\nu\not=0,\frac{d}{2}}$ has
conformal 
dimension $\frac{d}{2}+\nu$ and the field $\phi$ goes
to the border as
$x_{0}^{d/2-\nu}\partial_{n}\phi_0(\vec{x})$. Comparing
Eqs.(\ref{20},\ref{28}) we
notice that the normalizations of the boundary two-point
functions corresponding to the massive $\nu\not=0$ Dirichlet and Neumann
cases are in general different.

Now we consider the case $\nu=0\;$ that is $m^{2}=-\frac{d^{2}}{4}$.
Following the now usual steps we get
\beq
I_{N}^{\nu=0}=-\frac{\Gamma\left(
\frac{d}{2}
\right)}{d^{2}\pi^{\frac{d}{2}}}\;
\int d^{d}x \;
d^{d}y\;\partial_{n}\phi_{\epsilon}(\vec{x})
\;\partial_{n}\phi_{\epsilon}(\vec{y})\;\frac{\epsilon^{-d}}{ln^{2}
\epsilon}\;   
\frac{1}{|\;\vec{x}-\vec{y}\;|^{d}}\;+\;\cdots.
\label{580}
\eeq
Taking the limit
\beq
\lim _{\epsilon\rightarrow
0}\;( \epsilon^{\frac{d}{2}}\;ln\;\epsilon)^{-1} \;\partial_{n}
\phi_{\epsilon}(\vec{x})
= \partial_{n}\phi_{0}(\vec{x}),
\label{750}
\eeq
and making use of the AdS/CFT equivalence Eq.(\ref{27}) we find the
following boundary two-point function
\beq
\left<{\cal O}_{N}^{\nu=0}(\vec{x}){\cal
O}_{N}^{\nu=0}(\vec{y})\right>=
\frac{2\Gamma\left(\frac{d}{2}\right)}{d^{2}\pi^{\frac{d}{2}}}\;
\frac{1}{|\;
\vec{x}-\vec{y}\;|^{d}}\; .
\label{751}
\eeq
Then the conformal operator ${\cal O}_{N}^{\nu=0}$ on the boundary
CFT has conformal dimension $\frac{d}{2}$. Near the border the
scalar field has a logarithmic behavior $x_0^{d/2}\ln x_0 \;
\partial_{n}\phi_0(\vec{x})$. Again we find 
that the normalizations of the boundary
two-point functions corresponding to the $\nu=0$ Dirichlet and Neumann
cases are in general different.

\section{Mixed Boundary Condition}

Using the mixed boundary condition Eq.(\ref{100}) and again
Eqs.(\ref{4'},\ref{5'}) we get
\beq
a(\vec{k})=\frac{\epsilon^{-\frac{d}{2}}\;\psi_{\epsilon}(\vec{k})}
{[\beta\left(\alpha,\nu\right)+2\alpha\nu]K_{\nu}(k\epsilon)-\alpha
k\epsilon
K_{\nu+1}(k\epsilon)}\; ,
\label{1000}
\eeq
where $\beta(\alpha,\nu)$ is defined as 
\beq
\beta\left(\alpha,\nu\right)=1+\alpha\left( \frac{d}{2}-\nu\right).
\label{beta}
\eeq
Substituting Eq.(\ref{1000}) into Eq.(\ref{5'}) we find
\beq
\partial_{0}\phi_{\epsilon}(\vec{x})=\int
d^{d}y\;\psi_{\epsilon}(\vec{y})\int\frac{d^{d}k}{\left(
2\pi\right)^{d}}\;e^{-i\vec{k}\cdot\left(
\vec{x}-\vec{y}\right)}\;\epsilon^{-1}\;
\frac{\frac{d}{2}+\nu-
k\epsilon\;
\frac{K_{\nu+1}(k\epsilon)}
{K_{\nu}(k\epsilon)}}{\beta\left(\alpha,\nu\right)+2\alpha\nu-
\alpha k\epsilon\;\frac{K_{\nu+1}(k\epsilon)}{K_{\nu}(k\epsilon)}}\; .
\label{1001}
\eeq
Using this we can write the action Eq.(\ref{104}) as
\beq
I_{M}=-\frac{1}{2}\int d^{d}x\;
d^{d}y\;\psi_{\epsilon}(\vec{x})
\;\psi_{\epsilon}(\vec{y})\;\epsilon^{-d}\int\frac{d^{d}k}{\left(
2\pi\right)^{d}}\;e^{-i\vec{k}\cdot\left(
\vec{x}-\vec{y}\right)}\;
\frac{\frac{d}{2}+\nu-  
k\epsilon\;
\frac{K_{\nu+1}(k\epsilon)}
{K_{\nu}(k\epsilon)}}{\beta\left(\alpha,\nu\right)+2\alpha\nu-
\alpha k\epsilon\;\frac{K_{\nu+1}(k\epsilon)}{K_{\nu}(k\epsilon)}}\; .
\label{1002}
\eeq
As we shall see it is important to consider the cases $\beta=0$ and
$\beta\not=0$ separately in order to find out the relevant terms in
the series expansions of the Bessel functions.

Let us start with the case $\beta=0$. For $\beta=0$ we have
$\alpha=-1/(\frac{d}{2}-\nu)$ and $m\not=0$. We first consider the
massive case with 
$\nu\not=0,\frac{d}{2}$. Again we have to study separately the cases
with $\nu$ not integer and $\nu$ integer but non-zero. 
Let us first consider the case $\nu$ not integer. Making use of
Eq.(\ref{16}) with $\beta=0$ we get
\beq
\frac{d}{2} + \nu -
k\epsilon\;\frac{K_{\nu+1}(k\epsilon)}{K_{\nu}(k\epsilon)}=\frac{d}{2} -
\nu\; +\;\cdots, 
\label{2000}
\eeq
and
\beq
\frac{1}{\beta\left(\alpha,\nu\right)+2\alpha\nu-
\alpha
k\epsilon\;\frac{K_{\nu+1}(k\epsilon)}{K_{\nu}(k\epsilon)}}=
-\frac{\frac{d}{2}-\nu}{\frac{1}{2(1-\nu)}(k\epsilon)^{2}
-2^{1-2\nu}\frac{\Gamma(1-\nu)}{\Gamma(\nu)}(k\epsilon)^{2\nu}\;
+\;\cdots}\; .
\label{2001}
\eeq
Notice that for $0<\nu<1$ the dominating term in the denominator of
the r.h.s of Eq.(\ref{2001}) is
$(k\epsilon)^{2\nu}$. Substituting in Eq.(\ref{1002}) we get
\ba
I_{M}^{\beta=0,0<\nu<1}&=&
-2^{2\nu -2}\;\left(
\frac{d}{2}-\nu\right)^{2}
\;\frac{\Gamma(\nu)}{\Gamma(1-\nu)}\nonumber\\ &\times& \int
d^{d}x\; d^{d}y\;\psi_{\epsilon}(\vec{x})
\;\psi_{\epsilon}(\vec{y})\;\epsilon^{-2\nu-d}\;
\int\frac{d^{d}k}{\left(
2\pi\right)^{d}}\; e^{-i\vec{k}\cdot
(\vec{x}-\vec{y})}\;k^{-2\nu}\;+\;\cdots.\nonumber\\
\label{2002}
\ea
Integration over $\vec{k}$ thus yields
\beq
I_{M}^{\beta=0,0<\nu<1}=
-\frac{1}{4\pi^{\frac{d}{2}}}\;\left( \frac{d}{2}-\nu\right)^{2}
\;\frac{\Gamma(\frac{d}{2} - \nu)}{\Gamma(1-\nu)}
\; \int d^{d}x \;
d^{d}y\;\psi_{\epsilon}(\vec{x})
\;\psi_{\epsilon}(\vec{y})\;\frac{\epsilon^{-2(\nu+\frac{d}{2})}}{|\;
\vec{x}-\vec{y}\;|^{2(\frac{d}{2}-\nu)}}\; +\;\cdots.
\label{action1}
\eeq
For $\nu>1$ the dominating term in the denominator of the r.h.s of
Eq.(\ref{2001}) is $(k\epsilon)^{2}$ and Eq.(\ref{1002}) reads
\ba
I_{M}^{\beta=0,\nu>1}&=&
-(\nu-1)\;\left(
\frac{d}{2}-\nu\right)^{2}\nonumber\\ &\times&
\int 
d^{d}x\; d^{d}y\;\psi_{\epsilon}(\vec{x})
\;\psi_{\epsilon}(\vec{y})\;\epsilon^{-d-2}\;
\int\frac{d^{d}k}{\left(
2\pi\right)^{d}}\; e^{-i\vec{k}\cdot
(\vec{x}-\vec{y})}\;k^{-2}\;+\;\cdots.
\label{2003}
\ea
Integration over $\vec{k}$ is carried out for $d>2$ thus giving 
\beq
I_{M}^{\beta=0,\nu>1}=-(\nu-1)\left(
\frac{d}{2}-\nu\right)^{2}
\frac{\Gamma(\frac{d-2}{2})}{4\pi^{\frac{d}{2}}}\;
\int d^{d}x \;
d^{d}y\;\psi_{\epsilon}(\vec{x})  
\;\psi_{\epsilon}(\vec{y})\;
\frac{\epsilon^{-d-2}}{|\;
\vec{x}-\vec{y}\;|^{2\frac{d-2}{2}}}\; +\;\cdots.
\label{1003}
\eeq
For the case $\nu$ integer and non-zero  we make use of
Eq.(\ref{16'''}). The logarithmic terms vanish in the limit
$\epsilon\rightarrow 0$ and we find that the same result
Eq.(\ref{1003}) holds for $\nu$ integer and $\nu$ not integer.

Now in the action Eq.(\ref{action1}) we take the limit
\beq
\lim _{\epsilon\rightarrow
0}\epsilon^{-\nu-\frac{d}{2}}\psi_{\epsilon}(\vec{x})
= \psi_{0}(\vec{x})\; ,
\label{1004}
\eeq
whereas in the action Eq.(\ref{1003}) the limit to be taken is
\beq
\lim _{\epsilon\rightarrow
0}\epsilon^{-\frac{d}{2}-1}\psi_{\epsilon}(\vec{x})
= \psi_{0}(\vec{x}).
\label{1005}
\eeq
Using the AdS/CFT equivalence
\beq
\exp\left( -I _{AdS}\right) \equiv \left<\exp\left(\int d^{d}x \;
{\cal O}(\vec{x}) \; \psi_{0}(\vec{x})\right)\right>\; ,
\label{1006}
\eeq
we get the following boundary two-point functions
\beq
\left<{\cal O}_{M}^{\beta=0,0<\nu<1}(\vec{x}){\cal
O}_{M}^{\beta=0,0<\nu<1}(\vec{y})\right>=\frac{1}{2\pi^{\frac{d}{2}}}\;
\left( \frac{d}{2}-\nu\right)^{2}
\;\frac{\Gamma(\frac{d}{2}-\nu)}{\Gamma(1-\nu)}\;\frac{1}{|\;
\vec{x}-\vec{y}\;|^{2(\frac{d}{2}-\nu)}}\; ,
\label{1007}
\eeq
\beq
\left<{\cal O}_{M}^{\beta=0,\nu >1}(\vec{x}){\cal
O}_{M}^{\beta=0,\nu
>1}(\vec{y})\right>=
(\nu-1)\;\left( \frac{d}{2}-\nu\right)^{2}\;
\frac{\Gamma\left( \frac{d-2}{2}\right)}{2\pi^{\frac{d}{2}}}\;
\frac{1}{|\;
\vec{x}-\vec{y}\;|^{2\frac{d-2}{2}}}\; .
\label{1008}
\eeq
Then the operators ${\cal O}_{M}^{\beta=0,0<\nu<1}$ and ${\cal
O}_{M}^{\beta=0,\nu >1}$ have conformal dimensions
$\frac{d}{2}-\nu$ and $\frac{d-2}{2}$
respectively. For $0<\nu<1$ the field $\phi$ approaches the
boundary as $x_{0}^{d/2+\nu}\psi_0(\vec{x})$.  The 
derivation of the conformal dimension $\frac{d}{2}-\nu$ for its
associated boundary operator ${\cal O}_{M}^{\beta=0,0<\nu<1}$ is 
a rather striking feature. It is
worth noting that the upper constraint $\nu<1$ in Eq.(\ref{1007}) is
consistent with the unitarity bound.

For $\nu>1$ we found a boundary operator whose conformal dimension is
the unitarity bound $\frac{d-2}{2}$. Whereas we have already
found such a conformal dimension in the massless
Neumann case Eq.(\ref{31}) we have here a different normalization for the
boundary two-point function. We note that the
behavior of the scalar field for small $x_0$ is as it should be. 

Now we consider the case $\nu=0$, that is $m^{2}=-\frac{d^{2}}{4}$, 
keeping still $\alpha=-\frac{2}{d}$. We then find
\beq
I_{M}^{\beta=0,\nu=0}=-\frac{d^{2}\Gamma\left(   
\frac{d}{2}
\right)}{16\pi^{\frac{d}{2}}}\; \int d^{d}x \;
d^{d}y\;\psi_{\epsilon}(\vec{x})
\;\psi_{\epsilon}(\vec{y})\;
\frac{\epsilon^{-d}}{|\;
\vec{x}-\vec{y}\;|^{d}}\;+\;\cdots.
\label{1009}
\eeq 
Taking the limit
\beq
\lim _{\epsilon\rightarrow  
0}\;\epsilon^{-\frac{d}{2}}\psi_{\epsilon}(\vec{x})
= \psi_{0}(\vec{x}),
\label{1010}
\eeq
and making use of the AdS/CFT equivalence Eq.(\ref{1006}) we get the
following boundary two-point function
\beq
\left<{\cal O}_{M}^{\beta=0,\nu=0}(\vec{x}){\cal
O}_{M}^{\beta=0,\nu=0}(\vec{y})\right>=
\frac{d^{2}\Gamma\left(\frac{d}{2}\right)}{8\pi^{\frac{d}{2}}}\;
\frac{1}{|\;
\vec{x}-\vec{y}\;|^{d}}\; .
\label{1011}
\eeq
Then the conformal operator ${\cal O}_{M}^{\beta=0,\nu=0}$ on
the
boundary
CFT has conformal dimension $\frac{d}{2}$. Now the field $\phi$ goes
to the border as $x_0^{d/2} \psi_0(\vec{x})$ and no logarithm is
present. We find again that the normalization of the two-point function is
different from the corresponding ones of the Dirichlet and Neumann cases.

Let us now consider the case when $\beta \not=0$. We study first
the case $\nu\not=0$. Again the cases $\nu$ not integer and $\nu$
integer but non-zero must be considered separately. 
We first study the case $\nu$ not integer. Up to contact terms or higher
order terms in $\epsilon$ we find 
\beq
\frac{d}{2} + \nu -
k\epsilon\;\frac{K_{\nu+1}(k\epsilon)}{K_{\nu}(k\epsilon)}=
\left(\frac{d}{2} - \nu\right)\left[1-\frac{2^{1-2\nu}}{\frac{d}{2} -
\nu}\; \frac{\Gamma(1-\nu)}{\Gamma(\nu)}\;(k\epsilon)^{2\nu} \;
+\;\cdots\right],
\label{2005}
\eeq
and
\beq
\frac{1}{\beta\left(\alpha,\nu\right)+2\alpha\nu-
\alpha
k\epsilon\;\frac{K_{\nu+1}(k\epsilon)}{K_{\nu}(k\epsilon)}}=
\frac{1}{\beta(\alpha,\nu)}\left[1+\frac{2^{1-2\nu}\alpha}
{\beta(\alpha,\nu)}\;
\frac{\Gamma(1-\nu)}{\Gamma(\nu)}\;(k\epsilon)^{2\nu} \;
+\;\cdots\right].
\label{2006}
\eeq
Substituting in Eq.(\ref{1002}) we get
\ba
I_{M}^{\beta\not=0,\nu\not=0,\frac{d}{2}}&=&
\frac{1}{2^{2\nu}}\;\frac{1}{\beta^{2}(\alpha,\nu)}
\;\frac{\Gamma(1-\nu)}{\Gamma(\nu)}\nonumber\\ &\times& \int
d^{d}x\; d^{d}y\;\psi_{\epsilon}(\vec{x})
\;\psi_{\epsilon}(\vec{y})\;\epsilon^{2\nu-d}\;
\int\frac{d^{d}k}{\left(
2\pi\right)^{d}}\; e^{-i\vec{k}\cdot
(\vec{x}-\vec{y})}\;k^{2\nu}\;+\;\cdots.
\label{2007}
\ea
Integration over $\vec{k}$ yields 
\beq
I_{M}^{\beta\not=0,\nu\not=0,\frac{d}{2}}=-\frac{\nu}{\pi^{\frac{d}{2}}}\;
\frac{1}{\beta^{2}(\alpha,\nu)}\;\frac{\Gamma(\frac{d}{2}
+ \nu)}{\Gamma(\nu)}\;\int d^{d}x \;
d^{d}y\;\psi_{\epsilon}(\vec{x})
\;\psi_{\epsilon}(\vec{y})\; 
\frac{\epsilon^{2(\nu-\frac{d}{2})}}{|\;
\vec{x}-\vec{y}\;|^{2(\frac{d}{2}+\nu )}}\; +\cdots.
\label{action2}
\eeq
Consider now the case $\nu$ integer and non-zero. We find 
\beq
\frac{d}{2} + \nu -
k\epsilon\;\frac{K_{\nu+1}(k\epsilon)}{K_{\nu}(k\epsilon)}=
\left(\frac{d}{2} -
\nu\right)\left[1-(-1)^{\nu}\;\frac{2^{2-2\nu}}{\frac{d}{2} -
\nu}\; \frac{1}{\Gamma^{2}(\nu)}\;(k\epsilon)^{2\nu}\;ln\; k \;
+\;\cdots\right], 
\label{2008}
\eeq
and
\beq
\frac{1}{\beta\left(\alpha,\nu\right)+2\alpha\nu-
\alpha
k\epsilon\;\frac{K_{\nu+1}(k\epsilon)}{K_{\nu}(k\epsilon)}}=
\frac{1}{\beta(\alpha,\nu)}\left[1+(-1)^{\nu}\;\frac{2^{2-2\nu}\alpha}
{\beta(\alpha,\nu)}\;
\frac{1}{\Gamma^{2}(\nu)}\;(k\epsilon)^{2\nu}\;ln\; k \;
+\;\cdots\right].
\label{2009}
\eeq
Substituting in Eq.(\ref{1002}) we get
\ba
I_{M}^{\beta\not=0,\nu\not=0,\frac{d}{2}}
&=&(-1)^{\nu}\; 2^{1-2\nu}\;\frac{1}{\beta^{2}(\alpha,\nu)}
\;\frac{1}{\Gamma^{2}(\nu)}\nonumber\\ &\times& \int
d^{d}x\; d^{d}y\;\psi_{\epsilon}(\vec{x})
\;\psi_{\epsilon}(\vec{y})\;\epsilon^{2\nu-d}\;
\int\frac{d^{d}k}{\left(
2\pi\right)^{d}}\;e^{-i\vec{k}\cdot
(\vec{x}-\vec{y})}\;k^{2\nu}\;ln\; k \;
+\;\cdots.\nonumber\\
\label{2010}
\ea
Making use of Eq.(\ref{16'''''}) we get Eq.(\ref{action2}) again. So
both cases $\nu$ integer and $\nu$ not integer yield the same result. 

Now taking the limit
\beq
\lim _{\epsilon\rightarrow
0}\epsilon^{\nu-\frac{d}{2}}\psi_{\epsilon}(\vec{x})
= \psi_{0}(\vec{x}),
\label{1013}
\eeq
and making use of the AdS/CFT correspondence Eq.(\ref{1006}) we find the
following
boundary two-point function
\beq
\left<{\cal O}_{M}^{\beta\not=0,\nu\not=0,\frac{d}{2}}(\vec{x}){\cal
O}_{M}^{\beta\not=0,\nu\not=0,\frac{d}{2}}(\vec{y})\right>=
\frac{2\nu}{\pi^{\frac{d}{2}}}\;\frac{1}{\beta^{2}(\alpha,\nu)}\;
\frac{\Gamma(\frac{d}{2}+\nu)}{\Gamma(\nu)}\;\frac{1}{|\;
\vec{x}-\vec{y}\;|^{2(\frac{d}{2}+\nu)}}\; ,
\label{1014}
\eeq
so that the operator ${\cal
O}_{M}^{\beta\not=0,\nu\not=0,\frac{d}{2}}$ has conformal
dimension $\frac{d}{2}+\nu$.  From Eq.(\ref{1013}) we find that the
behavior of $\phi$ for 
small $x_0$ is as expected. Comparing Eqs.(\ref{20},\ref{28},\ref{1014})
we conclude that the normalizations of the boundary
two-point
functions corresponding to the massive $\nu\not=0$ Dirichlet, Neumann
and $\beta\not=0$ mixed cases are different. 

We now consider the case $\nu=0$ that is $m^{2}=-\frac{d^{2}}{4}$. 
We find  
\beq
I_{M}^{\beta\not=0,\nu=0}=-\frac{1}{\beta^{2}(\alpha,0)}
\frac{\Gamma\left(
\frac{d}{2}
\right)}{4\pi^{\frac{d}{2}}}\; \int d^{d}x \;
d^{d}y\;\psi_{\epsilon}(\vec{x})
\;\psi_{\epsilon}(\vec{y})\;\frac{\epsilon^{-d}}{ln^{2}\epsilon}\;
\frac{1}{|\;\vec{x}-\vec{y}\;|^{d}}\;+\;\cdots.
\label{1015}   
\eeq
Taking the limit
\beq
\lim _{\epsilon\rightarrow
0}\;( \epsilon^{\frac{d}{2}}\;ln\;\epsilon)^{-1} \;\psi_{\epsilon}(\vec{x})
= \psi_{0}(\vec{x}),
\label{1016}
\eeq
and making use of the AdS/CFT correspondence Eq.(\ref{1006}) we find the
following boundary two-point function
\beq
\left<{\cal O}_{M}^{\beta\not=0,\nu=0}(\vec{x}){\cal
O}_{M}^{\beta\not=0,\nu=0}(\vec{y})\right>=
\frac{1}{\beta^{2}(\alpha,0)}
\;\frac{\Gamma\left(\frac{d}{2}\right)}{2\pi^{\frac{d}{2}}}\;
\frac{1}{|\;
\vec{x}-\vec{y}\;|^{d}}\; ,  
\label{1017}
\eeq
so that the conformal operator ${\cal O}_{M}^{\beta\not=0,\nu=0}$ on the
boundary
CFT has conformal dimension $\frac{d}{2}$. For small $x_0$ we find a
logarithmic behavior $x_0^{d/2} \ln x_0 \;\psi_0(\vec{x})$. Again
the normalization of the two-point function is different when compared to
the corresponding ones of the Dirichlet, Neumann and $\beta=0$ mixed
cases.  

In the massless case we have $\nu=\frac{d}{2}$. For $d$ odd we make
use of
Eqs.(\ref{16},\ref{16'}), whereas for $d$ even we use
Eqs.(\ref{16'''},\ref{16'''''}). In both cases we get
\beq  
I_{M}^{\nu=\frac{d}{2}}
=-\frac{d}{2\pi^{\frac{d}{2}}}\;\frac{\Gamma(d)}{\Gamma(\frac{d}{2})}
\; \int d^{d}x \;
d^{d}y\;\psi_{\epsilon}(\vec{x})
\;\psi_{\epsilon}(\vec{y})\;\frac{1}{|\;
\vec{x}-\vec{y}\;|^{2d}}\;+\;\cdots.
\label{1018}
\eeq
Taking the limit
\beq
\lim _{\epsilon\rightarrow
0}\;\psi_{\epsilon}(\vec{x})
= \psi_{0}(\vec{x}),
\label{1019}
\eeq
and making use of the AdS/CFT equivalence Eq.(\ref{1006}) we get
the
following boundary two-point function
\beq
\left<{\cal O}_{M}^{\nu=\frac{d}{2}}(\vec{x}){\cal
O}_{M}^{\nu=\frac{d}{2}}(\vec{y})\right>=\frac{d}{\pi^{\frac{d}{2}}}\;
\frac{\Gamma(d)}{\Gamma(\frac{d}{2})}\;\frac{1}{|\;
\vec{x}-\vec{y}\;|^{2d}}\;.
\label{1020}
\eeq
Then the operator ${\cal O}_{M}^{\nu=\frac{d}{2}}$ has conformal
dimension $d$. The scalar field goes to the border as $\psi_0(\vec{x})$
as expected. Comparing Eqs.(\ref{21},\ref{1020}) we conclude that the
boundary
CFT's corresponding to the massless Dirichlet and mixed cases are equal.

\section{Conclusions}
We have shown how different boundary conditions in the AdS/CFT
correspondence allow us to derive boundary two-point functions which are
consistent with the unitarity bound. We have also done a careful
analysis of the particular cases when $\nu$ is an integer and we have
shown that when $\nu=0$ there are no double zeroes in the two-point
functions. 

In general the use of different
types of boundary conditions lead to boundary two-point functions
with different normalizations. It is not clear at this point whether
they are important or not. It is necessary to compute the
three-point functions in order to clarify if the different normalizations
are leading to different boundary CFT's.

In the Neumann case the unitarity bound is
obtained for $m=0$ while with mixed boundary conditions it is reached
when $\beta=0$ and $m^{2}>-d^{2}/4 + 1$. The corresponding two-point
functions have different normalizations. The conformal dimension $d/2 -
\nu$ is obtained in the case of mixed boundary condition with $\beta=0$
and $-d^{2}/4<m^{2}<-d^{2}/4 + 1$, and the normalization of
the corresponding boundary two-point function differs from the one found
in \cite{witten2}. 

We have also tried to relate our
formalism to the Legendre transform approach \cite{witten2}. We could
think that both formalisms are related through some field redefinition but
this is not the case. It is not possible to redefine the scalar field in
order to turn a Dirichlet boundary condition into a Neumann or mixed one.
If there is any relation between the two approaches it must be a very
subtle one.

Another important point is the interpretation of the new boundary
conditions in the string theory context. Dirichlet boundary conditions are
natural when thinking of the asymptotic behavior of the supergravity
fields reaching the border of the AdS space. Possibly Neumann and mixed
boundary conditions are related to more complex solutions involving
strings and membranes reaching the border in more subtle ways. This of
course needs a more detailed study.

\section{Acknowledgements}

P.M. acknowledges financial support by CAPES. V.O.R. 
is partially supported by CNPq and acknowledges a grant by FAPESP. 

\section{Appendix A. Boundary Two-Point Functions}
The coefficients $\nu$, $\alpha$ and $\beta(\alpha,\nu)$ are defined
in
Eqs.(\ref{5},\ref{100},\ref{beta}) respectively. Let us also define 
\beq
\sigma(\nu)=\frac{d}{2}-\nu\; .
\label{221}
\eeq
\subsection{Dirichlet Boundary Condition}
\ba
&&\left<{\cal O}_{D}^{\nu\not=0}(\vec{x}){\cal
O}_{D}^{\nu\not=0}(\vec{y})\right>=\frac{2\nu}{\pi^{\frac{d}{2}}}\;
\frac{\Gamma(\frac{d}{2}+\nu)}{\Gamma(\nu)}\;\frac{1}{|\;
\vec{x}-\vec{y}\;|^{2(\frac{d}{2}+\nu)}}
\label{260}\\
&&\left<{\cal O}_{D}^{\nu=\frac{d}{2}}(\vec{x}){\cal
O}_{D}^{\nu=\frac{d}{2}}(\vec{y})\right>=\frac{d}{\pi^{\frac{d}{2}}}\;
\frac{\Gamma(d)}{\Gamma(\frac{d}{2})}\;\frac{1}{|\;
\vec{x}-\vec{y}\;|^{2d}}
\label{261}\\
&&\left<{\cal O}_{D}^{\nu=0}(\vec{x}){\cal
O}_{D}^{\nu=0}(\vec{y})\right>=
\frac{\Gamma\left(\frac{d}{2}\right)}{2\pi^{\frac{d}{2}}}\;
\frac{1}{|\;
\vec{x}-\vec{y}\;|^{d}}
\ea
\label{262}

\subsection{Neumann Boundary Condition}
\ba
&&\left<{\cal O}_{N}^{\nu\not=0,\frac{d}{2}}(\vec{x}){\cal
O}_{N}^{\nu\not=0,\frac{d}{2}}(\vec{y})\right>=
\frac{1}{\sigma^{2}(\nu)}\;
\left<{\cal O}_{D}^{\nu\not=0}(\vec{x}){\cal
O}_{D}^{\nu\not=0}(\vec{y})\right>
\label{263}\\
&&\left<{\cal O}_{N}^{\nu=\frac{d}{2}}(\vec{x}){\cal
O}_{N}^{\nu=\frac{d}{2}}(\vec{y})\right>=
\frac{\Gamma(\frac{d}{2})}{2\pi^{\frac{d}{2}}}\;\frac{1}{|\;
\vec{x}-\vec{y}\;|^{2\frac{d-2}{2}}}
\label{264}\\
&&\left<{\cal O}_{N}^{\nu=0}(\vec{x}){\cal
O}_{N}^{\nu=0}(\vec{y})\right>=
\frac{1}{\sigma^{2}(0)}\;\left<{\cal O}_{D}^{\nu=0}(\vec{x}){\cal
O}_{D}^{\nu=0}(\vec{y})\right>
\label{265}
\ea
\subsection{Mixed Boundary Condition}
\ba
&&\left<{\cal O}_{M}^{\beta=0,0<\nu<1}(\vec{x}){\cal
O}_{M}^{\beta=0,0<\nu<1}(\vec{y})\right>=\sigma^{2}(\nu)
\frac{1}{2\pi^{\frac{d}{2}}}\;  
\;\frac{\Gamma(\frac{d}{2}-\nu)}{\Gamma(1-\nu)}\;\frac{1}{|\;
\vec{x}-\vec{y}\;|^{2(\frac{d}{2}-\nu)}}
\label{360}\\
&&\left<{\cal O}_{M}^{\beta=0,\nu >1}(\vec{x}){\cal
O}_{M}^{\beta=0,\nu >1}(\vec{y})\right>=\sigma^{2}(\nu)\;(\nu-1)\;
\frac{\Gamma\left( \frac{d-2}{2}\right)}{2\pi^{\frac{d}{2}}}\;
\frac{1}{|\;  
\vec{x}-\vec{y}\;|^{2\frac{d-2}{2}}}
\label{361}\\
&&\left<{\cal
O}_{M}^{\beta\not=0,\nu\not=0,\frac{d}{2}}(\vec{x}){\cal
O}_{M}^{\beta\not=0,\nu\not=0,\frac{d}{2}}(\vec{y})\right>=
\frac{1}{\beta^{2}(\alpha,\nu)}\;\left<{\cal O}_{D}^{\nu\not=0}(\vec{x}){\cal
O}_{D}^{\nu\not=0}(\vec{y})\right>
\label{719}\\
&&\left<{\cal O}_{M}^{\nu=\frac{d}{2}}(\vec{x}){\cal
O}_{M}^{\nu=\frac{d}{2}}(\vec{y})\right>=\left<{\cal
O}_{D}^{\nu=\frac{d}{2}}(\vec{x}){\cal
O}_{D}^{\nu=\frac{d}{2}}(\vec{y})\right>
\label{820}\\
&&\left<{\cal O}_{M}^{\beta\not=0,\nu=0}(\vec{x}){\cal
O}_{M}^{\beta\not=0,\nu=0}(\vec{y})\right>=\frac{1}{\beta^{2}(\alpha,0)}
\;\left<{\cal
O}_{D}^{\nu=0}(\vec{x}){\cal
O}_{D}^{\nu=0}(\vec{y})\right>
\label{721}\\
&&\left<{\cal O}_{M}^{\beta=0,\nu=0}(\vec{x}){\cal
O}_{M}^{\beta=0,\nu=0}(\vec{y})\right>=\sigma^{2}(0)
\;\left<{\cal
O}_{D}^{\nu=0}(\vec{x}){\cal
O}_{D}^{\nu=0}(\vec{y})\right>
\label{722}
\ea

\section{Appendix B. Some Useful Formulae}
\subsection{Series Expansions for the Modified Bessel Functions $K_{\nu}$}

For  $\nu$ not integer
\beq
K_{\nu}(z)=\frac{1}{2}\;\Gamma(\nu)\;\Gamma(1-\nu)\left(
\frac{z}{2}\right)^{-\nu}\left[ \sum_{n\geq 0}\frac{\left(
\frac{z}{2}\right)^{2n}}{n!\;\Gamma(n+1-\nu)}-\left(
\frac{z}{2}\right)^{2\nu}\sum_{n\geq
0}\frac{\left(
\frac{z}{2}\right)^{2n}}{n!\;\Gamma(n+1+\nu)}\right].
\label{16}
\eeq

For $\nu$ integer and non-zero

\ba
K_{\nu}(z)&=&\frac{1}{2}\;\left( \frac{z}{2}\right)^{-\nu}\;
\sum_{n=0}^{\nu-1}(-1)^{n}\;\frac{\Gamma( \nu - n )}{n!}
\left( \frac{z}{2}\right)^{2n}\nonumber\\
&-&(-1)^{\nu}\;\left(
\frac{z}{2}\right)^{\nu}\;\sum_{n\geq
0}\left[ln\left(\frac{z}{2}\right)-\frac{\lambda(n+1)+\lambda(\nu +
n + 1)}{2}\right]
\frac{\left(
\frac{z}{2}\right)^{2n}}{n!\;\Gamma(n+1+\nu)}\; ,\nonumber\\
\label{16'''}
\ea
where
\beq
\lambda(1)=-\gamma \qquad\qquad \lambda(n)=-\gamma +
\sum_{m=1}^{n-1}\frac{1}{m}
\qquad (n \geq 2),
\label{16''''}
\eeq
and $\gamma$ is the Euler constant.

For $\nu=0$
\beq
K_{0}(z)=
-\sum_{n\geq
0}\left[ln\left(\frac{z}{2}\right)-\lambda(n+1)\right]
\frac{\left(
\frac{z}{2}\right)^{2n}}{n!\;\Gamma(n+1)}\;.
\label{827}
\eeq
\subsection{Integration over the Momenta}
\ba
&&\int\frac{d^{d}k}{\left(
2\pi\right)^{d}}\; e^{-i\vec{k}\cdot\vec{x}}\;
k^{\rho}=C_{\rho}
\;\frac{1}{|\vec{x}|^{d+\rho}}\qquad\qquad\rho\not=-d,-d-2,...
\label{16'}\\   
&&\int\frac{d^{d}k}{\left(
2\pi\right)^{d}}\; e^{-i\vec{k}\cdot\vec{x}}\;
k^{\rho}\;ln\; k=\frac{dC_{\rho}}{d\rho}\;\frac{1}{|\vec{x}|^{d+\rho}} +
C_{\rho}\;\frac{ln|\vec{x}|}{|\vec{x}|^{d+\rho}}
\quad\rho\not=-d,-d-2,...,
\label{16'''''}
\ea
where
\beq
C_{\rho}=\frac{2^{\rho}}{\pi^{\frac{d}{2}}}\;
\frac{\Gamma(\frac{d+\rho}{2})}{\Gamma(-\frac{\rho}{2})}\;.
\label{16''}
\eeq

\end{document}